\documentclass[fleqn,usenatbib]{mnras}

\usepackage{newtxtext,newtxmath}

\usepackage[T1]{fontenc}
\usepackage{ae,aecompl}

\usepackage{graphicx}	
\usepackage{amsmath}	
\usepackage{amssymb}	
\usepackage{hyperref}
\usepackage{xcolor,color}

\definecolor{DT_red}{RGB}{200, 70, 80}

\newcommand{\GWAC}{GWAC }

\title[GWAC OT candidate vetting using CNN]{Vetting the optical transient candidates detected by the GWAC network using convolutional neural networks}

\author[D. Turpin et al.]{
Damien Turpin,$^{1,2}$\thanks{E-mail: damien.turpin@cea.fr}
M. Ganet,$^{3,4}$
S. Antier,$^{5,6}$
E. Bertin,$^{7}$
L. P. Xin,$^{1}$
N. Leroy,$^{5}$
C. Wu,$^{1}$
\newauthor{
Y. Xu,$^{1}$
X. H. Han,$^{1}$
H. B. Cai,$^{1}$
H. L. Li,$^{1}$
X. M. Lu$^{1}$
 and J. Y. Wei$^{1,8}$}
\\
$^{1}$CAS Key Laboratory of Space Astronomy and Technology, National Astronomical Observatories, Chinese Academy of Sciences, Beijing 100101, China\\
$^{2}$AIM, CEA, CNRS, Universit\'e Paris-Saclay, Universit\'e Paris Diderot, Sorbonne Paris Cit\'e, F-91191 Gif-sur-Yvette, France\\
$^{3}$ENSAE Paris, F-91120 Palaiseau, France\\
$^{4}$HEC Paris, F-78350 Jouy-en-Josas, France\\
$^{5}$LAL, Univ Paris-Sud, CNRS/IN2P3, Orsay, France\\
$^{6}$APC, Univ. Paris Diderot, CNRS/IN2P3, CEA/lrfu, Obs de Paris, Sorbonne Paris Cit\'e, France\\
$^{7}$UPMC-CNRS, UMR7095, Institut d'Astrophysique de Paris, 75014 Paris, France\\
$^{8}$School of Astronomy and Space Science, University of Chinese Academy of Sciences, 101408 Beijing, China
}

\date{Accepted XXX. Received YYY; in original form ZZZ}

\pubyear{2020}

\begin{document}
\label{firstpage}
\pagerange{\pageref{firstpage}--\pageref{lastpage}}
\maketitle

\begin{abstract}
The observation of the transient sky through a multitude of astrophysical messengers has led to several scientific breakthroughs these last two decades thanks to the fast evolution of the observational techniques and strategies employed by the astronomers. Now, it requires to be able to coordinate multi-wavelength and multi-messenger follow-up campaign with instruments both in space and on ground jointly capable of scanning a large fraction of the sky with a high imaging cadency and duty cycle. In the optical domain,
    the key challenge of the wide field of view telescopes covering tens to hundreds of square degrees is to deal with the detection, the identification and the classification of hundreds to thousands of optical transient (OT) candidates every night in a reasonable amount of time. In the last decade, new automated tools based on machine learning approaches have been developed to perform those tasks with a low computing time and a high classification efficiency. 
    In this paper, we present an efficient classification method using Convolutional Neural Networks (CNN) to discard any bogus falsely detected in astrophysical images in the optical domain. We designed this tool to improve the performances of the OT detection pipeline of the Ground Wide field Angle Cameras (GWAC) telescopes, a network of robotic telescopes aiming at monitoring the optical transient sky down to R=16 with a 15 seconds imaging cadency. 
    We applied our trained CNN classifier on a sample of 1472 GWAC OT candidates detected by the real-time detection pipeline. It yields a good classification performance with 94\% of well classified event and a false positive rate of 4\%.
\end{abstract}




\section{Introduction}

The time domain astronomy aim at studying transient phenomena having a wide variety of flux and time scales and detected with a very broad range of localization accuracies in the sky depending on the astrophysical messengers emitted (electromagnetic, gravitational waves and high-energy particles). For several centuries, the main observed transient phenomena were the supernovae (SNe) in the optical domain, tracing the violent fate of the most massive stars undergoing a core collapse or the thermonuclear explosion of white dwarfs accreting the matter of a companion star \citep[see for example][about the SNe classification]{GalYam17}. In the last century, the SNe were detected only at a rate of few per year\footnote{See for example \url{http://www.rochesterastronomy.org/snimages/snactive.html}}, mainly because the observational techniques and strategies were not optimized to frequently detect such rare events\footnote{The observed local (within 100 Mpc) supernovae rate is about 10$\mathrm{^{-4}~SNe\cdot yr^{-1}\cdot Mpc^{-3}}$ \citep{Horiuchi11}.}. Therefore, the workload pressure on the detection pipelines and classification procedures of those transients were easily manageable by involving human actions in several steps, especially knowing that SNe can be observed during several days to months after the initial explosion with a 1-meter class telescope.\\
A first major revolution in the transient sky astronomy came with the development of the high-energy x-ray and gamma-ray telescopes and the detection of new classes of transients such as the Gamma-ray bursts \citep[GRB;][]{Klebesadel73} or the flaring blazars \citep{Brown86, Robson88, Hartman92}. In addition to the high-energy emission, those transients also produce low energy broadband emission up to the radio wavelengths. Hence, multi-wavelength follow-up observations across the whole electromagnetic spectrum became crucial to get a global picture of the physical processes. The GRBs certainly represent one of the most extreme observational challenge for the follow-up telescopes as the short-living initial gamma-ray signal \citep[see the review on GRB physics by][]{Kumar15} can be very poorly localized within up to several tens of square degrees depending on the trigger instrument. Then, a race against time is engaged to catch the so-called multi-wavelength afterglow emission that is fading very quickly so that it usually becomes unreachable for a detection 1-2 days after the trigger time by any x-ray or optical facility. This kind of transient event has definitely led to the birth of a new type of astronomy where different type of electromagnetic facilities have to work together in near real-time to complete the scientific data sets. Two decades ago, in the optical domain, several groups started to develop networks of small aperture robotic telescopes (for example ROTSE, TAROT, BOOTES, MASTER) that were\footnote{Most of them are still in operation.} capable to respond to any alert and scan a large fraction of the night sky continuously with a high cadence \citep[][]{Marshall97,Akerlof03,Boer99,Klotz08,Castro-Tirado99,Lipunov10}. The multiplication of the synergies between the space and ground-based telescopes, all broadcasting alerts about a large variety of transient sources, has largely contributed to increase the flow of data to be analyzed on real-time (photometry, spectroscopy and polarimetry). \\
Currently, the increasing pressure on the data processing of the follow-up telescopes studying the transient sky is significantly accelerating with the recent birth of the multi-messenger (MM) astronomy adding the high energy-neutrinos (HEN) and gravitational wave (GW) events in the global alert broadcasting system. With the constant sensitivity improvements of the electromagnetic and the MM facilities, one can now regularly deal with the reception of several valuable transient alerts of any astrophysical type every night. In the next decade, the multiplication of the facilities dedicated to the study of the transient sky and being able to make an all-sky monitoring at even deeper sensitivities will continue to progress, e.g. the \textit{Large Synoptic Survey Telescope} (LSST) \citep{Ivezic08}, the \textit{Square Kilometer Array} (SKA) \citep{Taylor00}, KM3NeT \citep{Adrian-Martinez16}, SVOM \citep{Wei16} or the next generation of GW detectors LIGO/Virgo and Kagra \citep{Abbott18}. Those projects will definitely make the time-domain astronomy enter into the big data era. As an example, the LSST project \citep{Ivezic19} would produce 20 terabytes of data every night with the possibility of having several hundreds of thousands alerts per night starting from 2021 and running over ten years of operation. It should extend the known SNe catalog with more than three billions of new entries (more than two orders of magnitude in terms of detection rate compared to any current survey).\\
 In the optical domain several groups already developed synoptic surveys, like the Catalina Real-time Transient Survey \citep[CRTS;][]{Drake09}, PTF \citep{Law09}, ASAS-SN\footnote{\url{http://www.astronomy.ohio-state.edu/~assassin/index.shtml}}, PanSTARRS \citep{Chambers16}, ATLAS \citep{Tonry18}, ZTF \citep{Bellm19}, DES \citep{Goldstein15} or Gaia \citep{GaiaDR2}, that explore the transient sky in addition to their participation to the various multi-messenger follow-up campaigns \citep[see for example][]{Abbott17}. The data flow generated by those surveys are already no longer manageable in a reasonable amount of time by the standard techniques previously used for narrow field of view telescopes as shown for example for ZTF \citep{Mahabal19}. The standard transient detection pipelines were usually based on PSF-matching and the catalog cross-matching methods for the detection of new sources, followed by a human validation of each transient candidate for the classification task. The growing alert rates and data flows now force the astronomers to develop new observational strategies and techniques to quickly detect, identify and classify the numerous uncatalogued sources they catch every night in their extensive searches.\\  
New techniques using \textit{machine learning} algorithm are developed to perform robust automated classifications of hundreds up to thousands of sources every night in real-time. The classification task is usually split into two steps independently performed. First, the goal is to filter out the bogus sources from the real uncatalogued sources of interest \citep[e.g.][]{Masci17,Sanchez19,Mahabal19, Jia19} immediately after the detection. The second step goes deeper in the classification procedure by associating an astrophysical category to an identified transient based on its temporal and/or spectral properties \citep[e.g.][]{Morii16, Narayan18,Muthukrishna19}.
Among the zoo of \textit{machine learning} algorithms, convolutional neural networks (CNN) are now massively used for such tasks as they are well-adapted to ingest data containing multiple arrays like images \citep{Bishop06,LeCun15}. They employ multiple interconnected layers, similar to a neuronal network, to efficiently identify patterns in images and are therefore particularly suitable for the time-domain astronomy \citep{Gieseke17}.\\
In this paper, we investigate the possibility of using CNN for the vetting of the optical transient (OT) candidates that will be detected by the \textit{Ground Wide field Angle Cameras} network (GWAC).  The GWAC system is a synoptic optical survey which is currently able to instantaneously cover 2000 square degrees on the sky with a high imaging cadency of one frame every 15 seconds. In operation since 2017, GWAC is a part of the ground-based follow-up system of the SVOM mission \citep{Wei16}, the next generation of space mission dedicated to the study of the multi-wavelength transient sky. It already provides a large data flow that must be smoothly digested by the real-time data processing pipeline as well as a significant amount of OT candidates sometimes well identified as real transients such as dwarf novae outbursts recently discovered in the GWAC survey \citep{Wang19}. 
The GWAC network is a perfect example of the evolution of the optical facilities that emerge nowadays to study the transient phenomena. It brings new observational challenges  which have to be solved in order to exploit the full capabilities of the instruments.
\\
In the section \ref{sec:GWAC_tel}, we will describe the GWAC system and the transient detection pipeline which is currently running. Then, in the section \ref{sec:classifier}, we will introduce the deep machine learning classifier we set up for the vetting of the GWAC OT. The classification results and performances will be presented in the section \ref{sec:results} and we finally draw our conclusions and perspectives for this work in the section \ref{sec:conclusion}.

\section{The GWAC telescopes}
\label{sec:GWAC_tel}
\subsection{Instrumentation setup}
\begin{figure}
	\centering
	\includegraphics[trim = 0 0 160 30,clip=true, width=0.7\columnwidth]{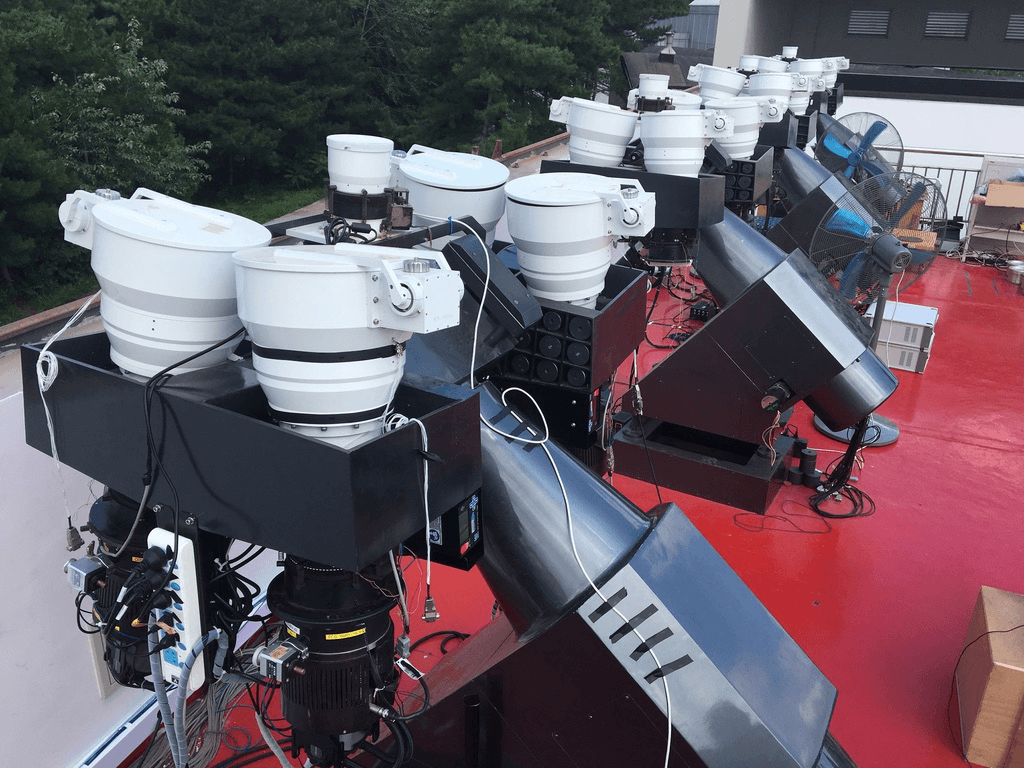}
	\caption{The GWAC telescope network at the Xinglong observatory in China. Currently, 4 mounts are operational among 10 at completion. Each mount is equipped with four JFoV camera (18 cm) and one FFoV camera (3.5 cm) located at the center of the mount. The total FoV of the current GWAC network is about 2000 sq.deg.}
	\label{Fig:gwac}
\end{figure}

Since the end of 2017, the Ground Wide field Angle Cameras telescopes are under development in China at the Xinglong Observatory. 
Each \GWAC{}telescope mount is equipped with five cameras: four JFoV cameras (4k $\times$ 4k CCD E2V camera with an aperture of 180 mm) and 1 FFoV camera (3k $\times$ 3k CCD camera with an aperture of 35 mm) used to monitor the sky seeing and brightness conditions, see Figure \ref{Fig:gwac}. The main scientific instruments, the JFoV cameras, cover a field of view of about $12.4^\circ \times 12.4^\circ$ per camera ($\sim$150 square degrees per camera). Taking into account the overlaps between the fields of view of the 4 JFoV cameras, a \GWAC{} mount finally covers 500 square degrees on the sky. Each JFoV camera is designed to reach an unfiltered limiting magnitude of about 16 in a dark night for 10 seconds of exposure. A stacking analysis of the single frames can be performed on real-time to reach a maximum limiting magnitude of R$\sim$18 in clear and dark night as shown in \citep{GWACO2}. 
\begin{table}
		\caption{\textcolor{black}{Characteristics of the  \GWAC{} JFoV cameras.}}
		\label{tab:GWAC_JFoV}
		\begin{center}
		\begin{tabular}{ll}
			\hline\hline\noalign{\smallskip}
			Parameter & value \\
			\hline
			Field of view & 150 sq.deg. \\
			diameter & 18 cm \\
			CCD pixel size & 13 $\mu$m \\
			Pixel scale & 11.7 arcsec \\
			Readout noise & 14 e$^-$ \\
			$\mathrm{R_{lim}}$ (single 10s/stack frame 1.5h) & 16/18 \\ 
			\noalign{\smallskip}\hline
		\end{tabular}
		\end{center}
	\end{table}
\subsection{The GWAC optical transient detection pipeline}
\label{sec:GWAC_pipeline}
\begin{figure*}
	\includegraphics[trim = 0 0 0 0,clip=true, width=1.0\columnwidth]{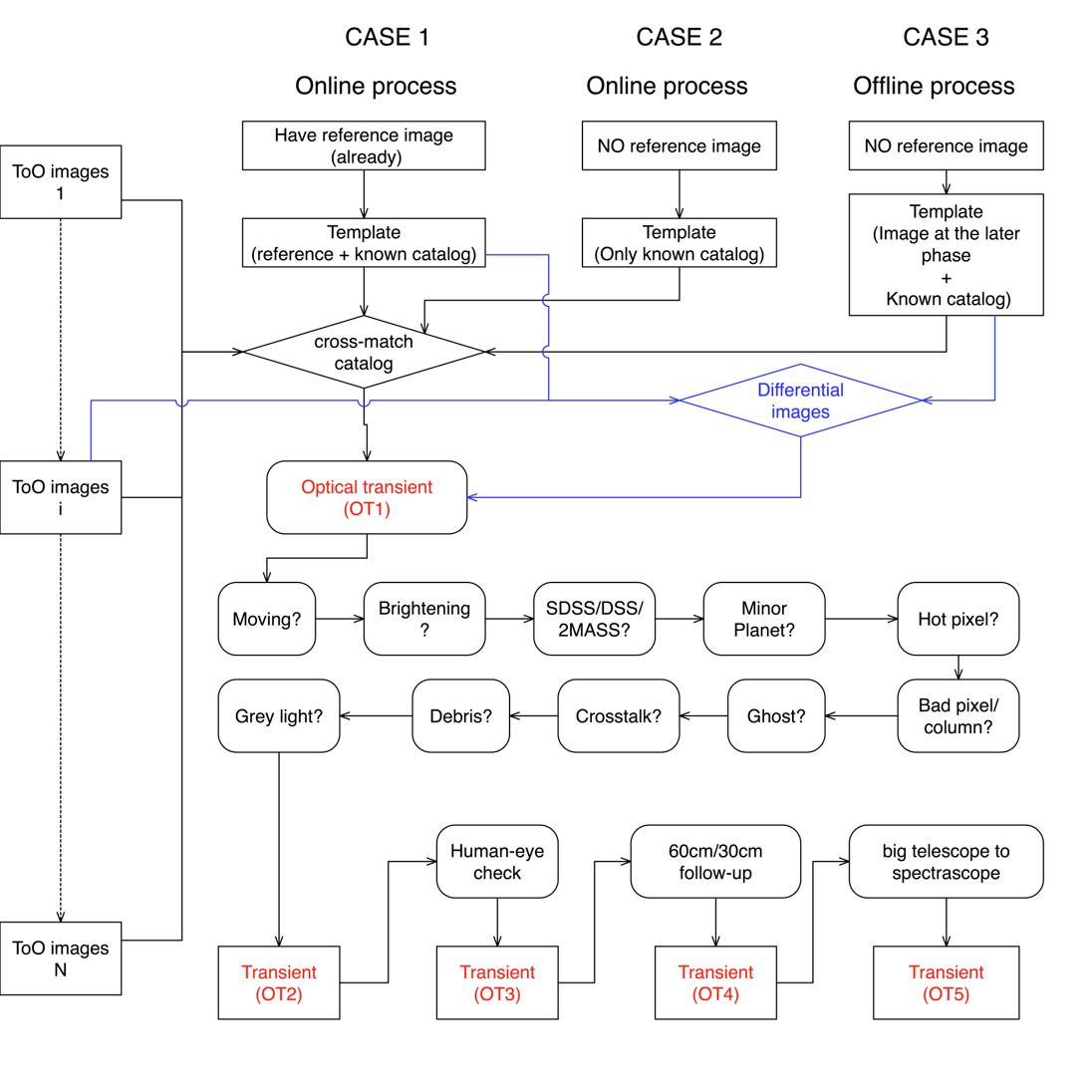}
	\caption{A schematic view of the current GWAC detection pipeline setup to detect and identify the optical transient sources in both single and stacked images.}
	\label{fig:det_pipeline}
\end{figure*}

The search for OT in GWAC data is made through several steps from the detection of candidates to their identification as being real variable/transient sources. 
The raw images are first pre-processed camera per camera to correct them from the Dark and the Bias offsets and to make the WCS (World Coordinates System) calibration. Those calibrated images are then automatically and independently analyzed by two pipelines to search for OT candidates. These two pipelines make use of standard methods comparing the scientific images with reference images taken much earlier such as the catalog cross-matching and the differential image analysis (DIA). Concerning the GWAC system more details can be found in \citep{GWACO2,Wang19} but typically a new source is detected once it fulfills the following criteria:
\begin{enumerate}
    \item The source has a signal-to-noise (SNR) ratio $\ge$ 5 and is not detected down a SNR = 5 in the reference images
    \item The source is detected in several successive images
    \item The point spread function (PSF) of the source shall be stellar-like profile, i.e. a 2D gaussian profile.
    \item no any CCD defect is detected in a region of 6 pixels around the source.
\end{enumerate}
The uncatalogued sources extracted from those analysis form the preliminary OT candidate list named \textit{OT1 candidates}. Then, several filters are applied on the source candidate parameters (the Full Width at Half Maximum -FWHM-, the SNR, the optical peak flux, the source position, etc.) on at least 5 successive images. Practically speaking, these filters aim to clean the \textit{OT1 candidates} from most of the spurious sources like the hot pixels or cosmic ray tracks. If at least 2/5 images pass the selection criteria, the OT candidates is kept otherwise it is rejected. A catalog cross matching filter using deeper catalogs is then applied to the \textit{OT1 candidates} that passed the first selection criteria, see Figure \ref{fig:det_pipeline}. Catalogs such as Gaia DR2 \citep{GaiaDR2}, PanStarrs DR1 \citep{Chambers16}, 2MASS \citep{Skrutskie06}, Galex DR5 \citep{Galex_DR5} or public databases on solar system objects such as the Minor Planet center\footnote{\url{https://minorplanetcenter.net/iau/mpc.html}} are used to perform this task.
 
After passing all of those filters, the remaining candidates are grouped in the \textit{OT2 candidates}. Sub-images are then cropped from each initial 4k $\times$ 4k JFoV images and subtracted from the sky background contribution to make 100 $\times$ 100 pixel-sized finding charts centered at the positions of each selected \textit{OT2 candidate}. These finding charts are then checked one by one by a human eyed-check analysis. Simultaneously, two 60 cm robotic telescopes (GWAC-F60A and GWAC-F60B) located beside the GWAC telescopes at the Xinglong Observatory automatically perform follow-up observations of any source found by the GWAC system in order to help the GWAC scientist on duty to finally confirm the genuineness of a given \textit{OT2 candidate}. Once the OT candidates are confirmed as being real transient sources, additional follow-up observations can be triggered with larger telescopes and public alerts can be released. This kind of detection pipeline is commonly used in the time domain astronomy. However, while it is robust enough for telescopes with a very limited field of view (typically few tens of arcminute), it turns to be no longer the optimal solution for telescopes covering hundreds of square degrees in the sky like the GWAC system as explained in the next section.

\subsection{Data flow and false detection rate}
The GWAC telescope network is operated in a sky survey mode following a pre-defined sky grid pointing strategy searching for bright optical transient events with a  minimum of sub-minute time scale. Since the beginning of 2019, four \GWAC{}mounts (16 JFoV cameras) are operational but at completion, the full system will be composed of 10 mounts. 
This setup implies the collection of a huge amount of data every night with typically between 6000-8000 images taken each night for a single telescope mount. When the observational conditions are optimum, the current network can generate as a whole as many as 24 000 images per night (up to 80k images per night for the complete network).\\ 
When using the detection pipeline described above and in Figure \ref{fig:det_pipeline}, the difficulties encountered with the GWAC telescopes system mainly come from the data flow and subsequently, the large false detection rate it can produce. The data flow is generated by the image cadency (the exposure time) and the number of operated cameras. The false detection rate is partly due to the data flow itself but it is also strongly dependent on the optical sensitivity of the instruments, their field of views and the strictness of the transient selection criteria. In addition, the large field of view of the single GWAC cameras (150 sq$\cdot$deg) combined with the limited size of the CCD detectors 
produces a large pixel scale of $\rm{11.7~arcsec\cdot pix^{-1}}$ and image distortion effects (while corrected in our images). These two factors make the use of the catalog cross matching and the differential image analysis even more complicated. This usually results in the production of additional fake detections populating the \textit{OT1 candidates} category. While the standard filtering algorithms are able to clean many fake \textit{OT1 candidates}, there is still a large fraction of them that pass through the filters. Typically, for one GWAC telescope mount, the number of \textit{OT1 candidates} can be as numerous as several hundreds in a single night depending on the observational conditions. Our standard filtering algorithms then reduces this number to several tens up to a few hundreds. Those ones then must be manually vetted both by humans and further follow-up observations. \\
Multiplying this task to the number of GWAC mounts and one can easily understand that this "true or false" classification task becomes no longer manageable both by the GWAC-F60 follow-up telescopes and the GWAC scientists in a reasonable amount of time. Therefore, our GWAC-F60 telescopes can be rapidly unable to ingest the quantity of triggers and additionally they can no longer smoothly follow their own observation plans independently of the GWAC camera activities. Moreover, the increase in our duty scientists workload finally make them no longer being able to focus their efforts on the most promising events. The identification and classification processes of a genuine transient source then undergo a long delay which is not compatible with the scientific purposes of the GWAC system that aim to quickly identify short-lived optical transient sources.

\section{A deep learning classifier}
\label{sec:classifier}

Our goal is to improve the current detection pipeline of the GWAC system, especially in easing the \textit{OT1 candidates} classification and making the human decision-taking process more responsive. As shown previously, there is a crucial need for a classification that distinguishes the astrophysical sources from the GWAC alert stream prior to build the \textit{OT2 candidates} list. 
Before going deeper into the details, we start to define few acronyms that we will use all along the paper:
\begin{itemize}
    \item \textit{ROS}: real optical sources in an image.
    \item \textit{FOS}: fake optical sources in an image.
    \item \textit{ROT}: real optical transients. A ROT is actually a ROS present in a series of images and showing a significant flux variation.
    \item \textit{FOT}: fake optical transients.
    \item \textit{TP}: true positives, i.e. the OT candidates well classified as \textit{ROT} or \textit{ROS}.
    \item \textit{TN}: true negatives, i.e. the OT candidates well classified as \textit{FOT} or \textit{FOS}.
    \item \textit{FP}: false positives, i.e. the OT candidates classified as \textit{ROT} or \textit{ROS} while there are actually \textit{FOT} or \textit{FOS}.
    \item \textit{FN}: false negatives, i.e. the OT candidates classified as \textit{FOT} or \textit{FOS} while there are actually \textit{ROT} or \textit{ROS}.
\end{itemize}
One immediately understands that our classifier must minimize the number of \textit{FP} and \textit{FN} to limit the contamination of the \textit{OT2 candidates} sample by any bogus in one hand and to avoid too many losses of \textit{ROT} because of misclassifications in the other hand. The final goal is to obtain a classification accuracy greater than 90\% with a \textit{FN} classification not as great as 2\%. Indeed, we prefer to keep more false positives (FP) instead of losing too many transients falsely classified as bogus (FN) in the classification process. To perform this task, we used a Convolutional Neural Networks algorithm. This choice is firstly motivated by the fact that CNNs are very well adapted for pattern recognition in images \citep{Bishop06, LeCun15} and have been already robustly tested with success for many different purposes in astronomy \citep{Bloom12,Calleja04,duBuisson15,Mahabal19}. Secondly, the CNNs have demonstrated excellent classification performances compared to other standard and deep \textit{machine learning} methods with a minimum of implementation \citep{Gieseke17}.

\subsection{The CNN model architecture and implementation}
\label{sec:CNN_implementation}
While this kind of "true or False" classification game does not require in principle a very deep and complex network 
structure, a too basic network may also have limited performances even considering such a "simple" task as noticed by \cite{Gieseke17}.
We therefore built a CNN code using an architecture composed of two convolutional layers 
, two pooling layers, one ReLu and one softmax hidden layers, see the details in Table \ref{tab:CNN_archi}. The pooling layers were kept to $2 \times 2 $ bin size due to the small size of some objects projected in the large GWAC pixel scale. The cross-entropy function was used as a loss function to give a high weight for very confident false positives, which we strongly want to avoid.
\begin{table}
  \begin{center}
    \caption{The CNN structure used in this work.}
    \label{tab:CNN_archi}
    \begin{tabular}{l|c|l} 
      \textbf{Layers} & \textbf{Sizes}& \textbf{Characteristics} \\
      \hline
      convolution & 32 $\times$ 32 & $3 \times 3$ kernels\\  
      & & activation:relu \\
      max pooling &  32 $\times$ 32  & $2 \times 2$ \\
      convolution &  32 $\times$ 32 & $3 \times 3$ kernels\\  
      & & activation:relu \\
      max pooling &  32 $\times$ 32 &$2 \times 2$\\
      dense & 15 & activation:relu\\
      dense & 2 & activation:Softmax \\
    \end{tabular}
  \end{center}
\end{table}
The CNN was implemented in \textit{Python v3.6}, using the \textit{Keras}\footnote{\url{https://keras.io/}. See also \citep{Geron} for a review of the usages of \textit{Keras}.} package with \textit{TensorFlow2}\footnote{\url{https://www.tensorflow.org/}\\ \url{https://github.com/tensorflow/tensorflow}}. The \textit{Keras} package has the advantages to provide built-in diagnostic tools and a compact code writing which allow for a relative ease of use.
A \textit{Keras} Adam optimizer was used with a low learning rate (lr=0.0001) after witnessing disappointing convergence properties.\\

\begin{figure*}
\centering
\includegraphics[width=0.8\textwidth]{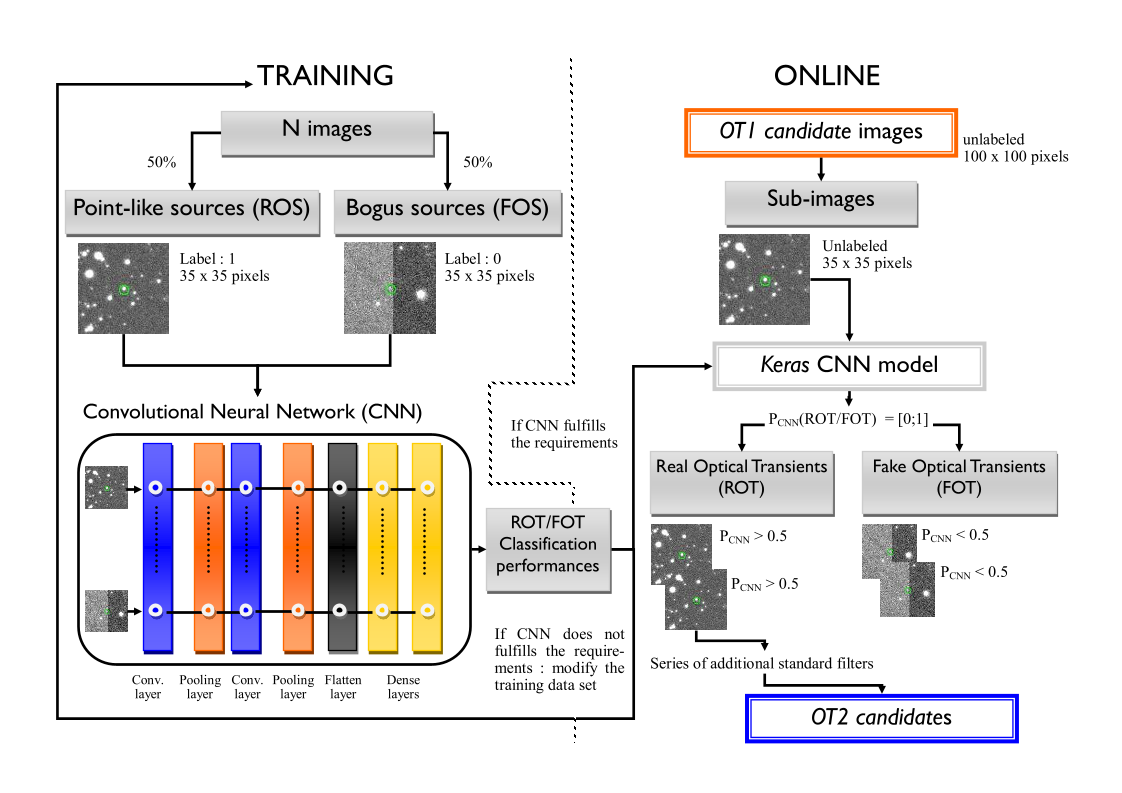}
\caption{Schematic view of the implementation of the classifier tool from the training of the CNN algorithm (\textit{left side}) to the use of the CNN \textit{Keras} model to make the vetting of the GWAC \textit{OT1 candidates} on real-time. We used a large data set of N = 200 000 images to train the CNN.}
\label{fig:CNNimple}
\end{figure*}

As an input, our CNN algorithm uses background-subtracted finding charts (100 $\times$ 100 pixels) of the \textit{OT1 candidates}. We then select only the central part of those images (35 $\times$ 35 pixels) for the classification. This choice is motivated to have a high learning rate as the CNN requires to be trained on an extensive amount of data (typically of the order of a minimum of 10$^5$ images) while keeping enough informations (background and a minimum number of sources) in the sub-images for the pattern recognition.
Before being able to give any classification on our OT candidates, the CNN must be trained to recognize patterns in our images. When a CNN layer receives an input, an output is then produced to feed the next layer. As long as the inputs are transformed into outputs, a series of several weights is produced to finally converge and build a final probabilistic rank between 0 and 1. The training phase contains several epochs of test to make the final convergence. The CNN ranking is then compared to the image labeling previously made by our expert scientists which consists in giving either a mark "1" to sub-images containing a \textit{ROT} or "0" if they contain \textit{FOT}. Therefore, this comparison method gives an idea on the level of agreement or disagreement of the CNN decision with the human classification. If a disagreement is frequently observed, it means either the CNN architecture is not optimised for our classification purpose or the human labeling is not correct. In such case, the CNN architecture and/or the labeling has to be revised until a good agreement is found. \\
At the end of the training, we  build a \textit{Keras} python model of our CNN to be used later to classify any OT candidate detected by the GWAC real-time detection pipeline.
Before using the CNN model in production, a final test of the classification performances is usually performed on an image sample that has never been used previously. If the CNN model reaches the classification requirements, i.e. if the number of misclassified sources is consistent with our scientific requirements, it can then be used to classify the genuineness of any GWAC optical transient candidates. On the contrary, if the classification is not good enough, a new training with a data set more representative of the GWAC OT candidate images must be performed until the classification requirements are fulfilled. \\
We illustrate, in figure \ref{fig:CNNimple}, the implementation we set up for both the training and the validation of the CNN algorithm as well as how it should be inserted in the detection pipeline of GWAC during real-time data taking.\\

\subsection{Training data set}
The classification of the sources into different astrophysical categories can be challenging. Indeed, transient sources are rare events and one might not have collected enough images of transient sources for the training. Some techniques can be used for the augmentation of the training data set such as simulating images of transient sources with a physical or empirical model or adding rotated images of real transient sources which artificially produces a new background and source distribution compared to the initial images as suggested by \cite{Gieseke17}. Typically, several tens of thousands images are needed to obtain a well trained CNN model.\\
Classifying our detected transients into several astrophysical categories based on additional informations such as the spectral and flux time evolution is actually beyond the scope of this work. For our purpose, our bogus/real source classification tasks is independent of the nature of the transient as long as it is supposed to be a point-like source in the images. As a consequence, we avoid the problem of having too few images of real GWAC transients to train the CNN. Instead, we can directly extract point-like sources in GWAC images to build our sample of \textit{ROS} images.
Our training data set is finally composed of 200 000 sub-images (35 $\times$ 35 pixels) with an equal distribution between \textit{FOS} and \textit{ROS}. Among them, 180 000 are directly used to train the algorithm while the 20 000 remaining images are used to validate each training epoch.\\ 
\subsubsection{Details on the \textit{ROS} image sample}
\begin{figure}
\centering
\includegraphics[width=0.35\columnwidth]{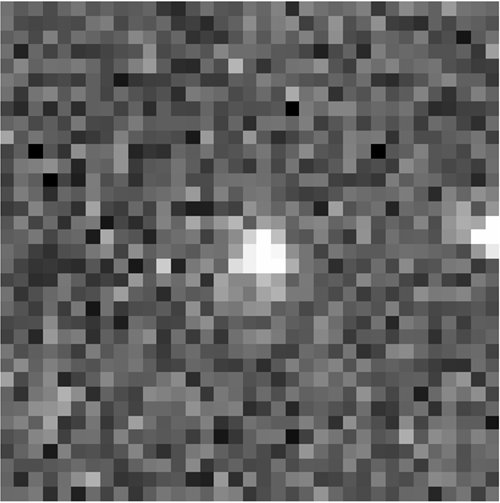}
\includegraphics[width=0.35\columnwidth]{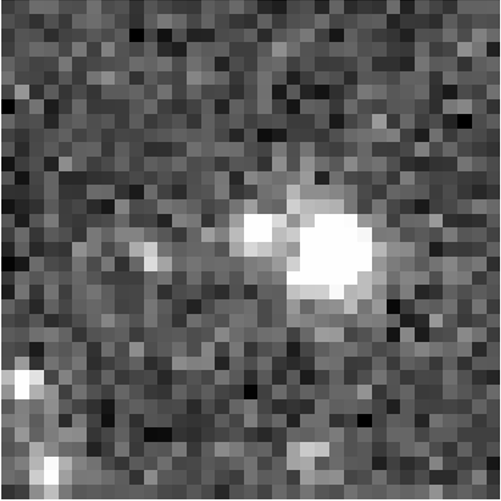}
\includegraphics[width=0.35\columnwidth]{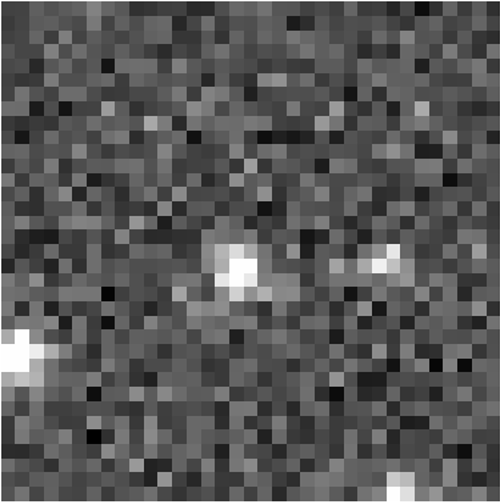}
\includegraphics[width=0.35\columnwidth]{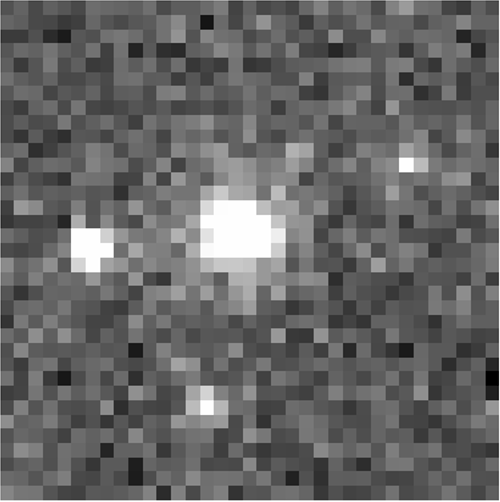}
\includegraphics[width=0.35\columnwidth]{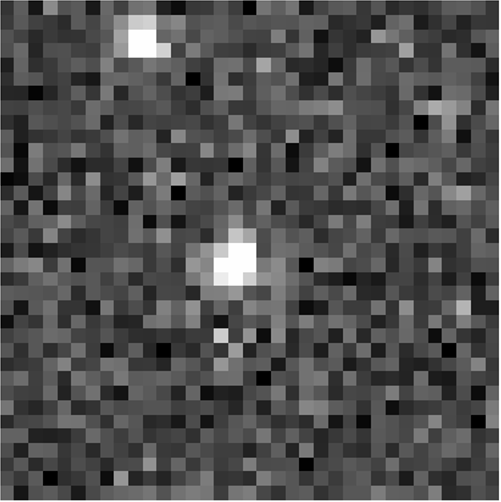}
\includegraphics[width=0.35\columnwidth]{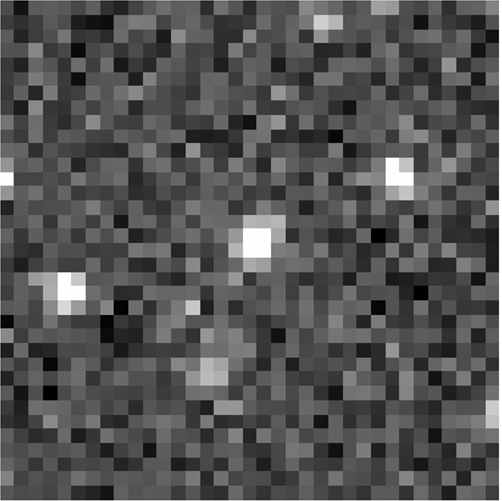}
\caption{Example of some sub-images (35$\times$ 35 pixels) centered ($\pm$ 2 pixel) at the position of point-like sources (ROTs) extracted from the GWAC 4k $\times$ 4k images. 100 000 images similar to these ones are produced to build our ROT training data set. Note that we extract point-like source with no prescription relative to their position in the original image (close to the edge or not, located in a dense star field, etc.).}
\label{fig:ROT_sample}
\end{figure}

The \textit{ROS} sub-image sample is built from several 4k $\times$ 4k GWAC images taken from the same camera during one year of operation. Therefore, we have at our disposal a complete overview of the observational and sky background conditions we can encounter at the GWAC site. The 4k $\times$ 4k initial images are chosen randomly and background-subtracted to follow the GWAC detection pipeline process previously described in section \ref{sec:GWAC_pipeline}. In each of the selected images, we extracted the position of the point sources detected by the \textit{Sextractor} software \citep{Bertin96} at the 3$\sigma$ confidence level. From this list of sources, we then randomly cropped 35 $\times$ 35 pixels sub-images around the \textit{Sextractor} positions of 100 randomly chosen sources. However, we make a selection cut on the instrumental magnitudes estimated by \textit{Sextractor} as we want to avoid very bright or "saturated" stars that may produce artefacts such as blooming effect. During the source extraction process and the creation of the finding charts, we noticed that the current GWAC detection pipeline can sometimes shift the centroid of the OT candidate from the center of the finding charts from 1 pixel at maximum in any direction. To be as close as possible to the GWAC pipeline output, we also reproduced this trend for each of our \textit{ROS} sub-image. The centroid of each extracted \textit{ROS} is therefore shifted in position in all the direction possible by an increment (uniformly) randomly chosen in the range [-1;1] pixel. We reproduced this operation on 1000 different 4k $\times$ 4k images to obtain a final sample of 100 000 images of \textit{ROS}. We show, in figure \ref{fig:ROT_sample}, a sub-sample of \textit{ROS} images we used for the training of our CNN.
\subsubsection{Details on the \textit{FOS} image sample}

While the \textit{ROS} should all have a similar 2D gaussian profile (in the ideal case with negligible distortion effects) it is no longer the case for \textit{FOS}. Indeed, a large variety of bogus can lead to false detections such as cosmic-ray tracks, hot pixels, bad  pixels, crosstalks artefacts, dusts, irregularities in the sky background contribution, etc. Therefore, our \textit{FOS} training data set must reproduce as close as possible such bogus shape distribution.\\ 
As it is actually very complicated to exactly mimic all the types of bogus we may encounter, we finally divided our bogus in several categories that are easily reproducible and correspond to the most frequent type of the bogus we encounter in GWAC images: \textit{hot pixels}, \textit{background noise}, \textit{bad column of pixels}, \textit{dark pixels} and a sky background with a significant \textit{light gradient}. To reach the same statistics than the \textit{ROS} training sample we had to use data augmentation techniques as we did not get enough images of all the categories of bogus. We simulated 100 000 images of bogus (50\% of the full training data set) in equal proportions between our five categories defined above. Our bogus simulator starts with the same process than for extracting \textit{ROS} from the 4k $\times$ 4k GWAC images. From the background-subtracted initial images, we extract 35 $\times$ 35 pixel sub-images and add a bogus in the central position similarly as we did for the \textit{ROS}. Then comes the difference in the process, depending on the bogus to simulate we crop different parts of the 4k $\times$ 4k images according to the following criteria:
\begin{enumerate}
    \item For the \textit{hot pixel} sub-images: no \textit{Sextractor} sources should have a position (X,Y) in the sub-image consistent within a region of at least 6 pixels around the central pixel (X0,Y0): $(X-X0)^{2} + (Y-Y0)^{2} \ge 36$
    \item For the \textit{noisy} sub-images: no \textit{Sextractor} sources should be present in the sub-images only the background residual noise.
    \item For the \textit{bad pixel columns} sub-images: any sub-image randomly chosen is suitable.
    \item For the \textit{dark pixels} sub-images: any sub-image randomly chosen is suitable.
    \item For the \textit{non uniform sky background} sub-images: we select only the position of brightest stars (estimated by \textit{Sextractor}), even the saturated stars, that produce a light gradient in the surrounding pixels. The distance from the center of the image to the position of bright star centroid can span from $d \in [6;15]$ pixels.
\end{enumerate}

\begin{figure}
\hspace{-0.5cm}\textbf{(a) simulated bogus} ~~~~~~~~~~~~~~~ \textbf{(b) observed bogus}\\
(35 $\times$ 35 pix) ~~~~~~~~~~~~~~~~~~~~~~~ (100 $\times$ 100 pix)\\
\centering
\textit{hot pixels}\\
\includegraphics[trim = 0 0 0 0,clip=true, width=0.45\columnwidth]{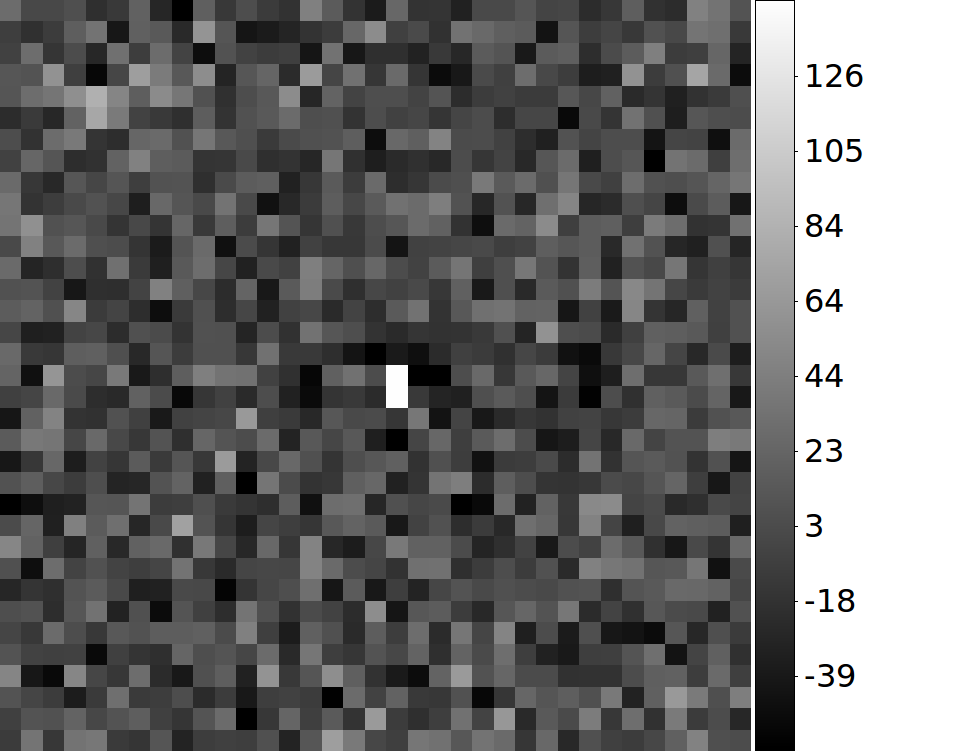}
\includegraphics[trim = 0 0 0 0,clip=true, width=0.45\columnwidth]{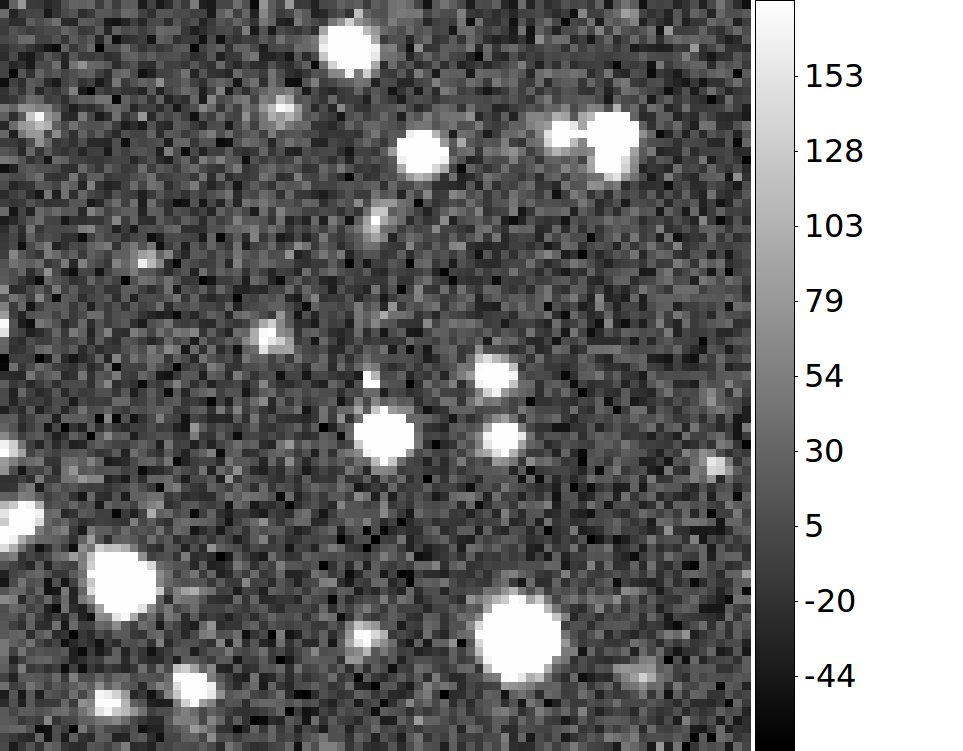}\\
\textit{background noise}\\
\includegraphics[trim = 0 0 0 0,clip=true, width=0.45\columnwidth]{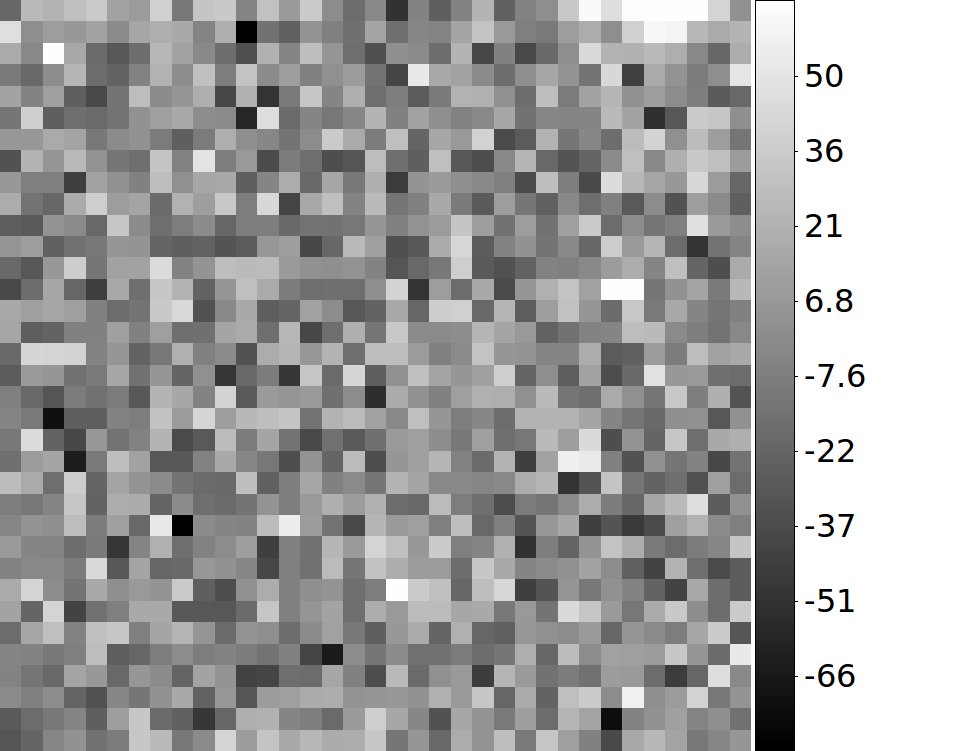}
\includegraphics[trim = 0 0 0 0,clip=true, width=0.45\columnwidth]{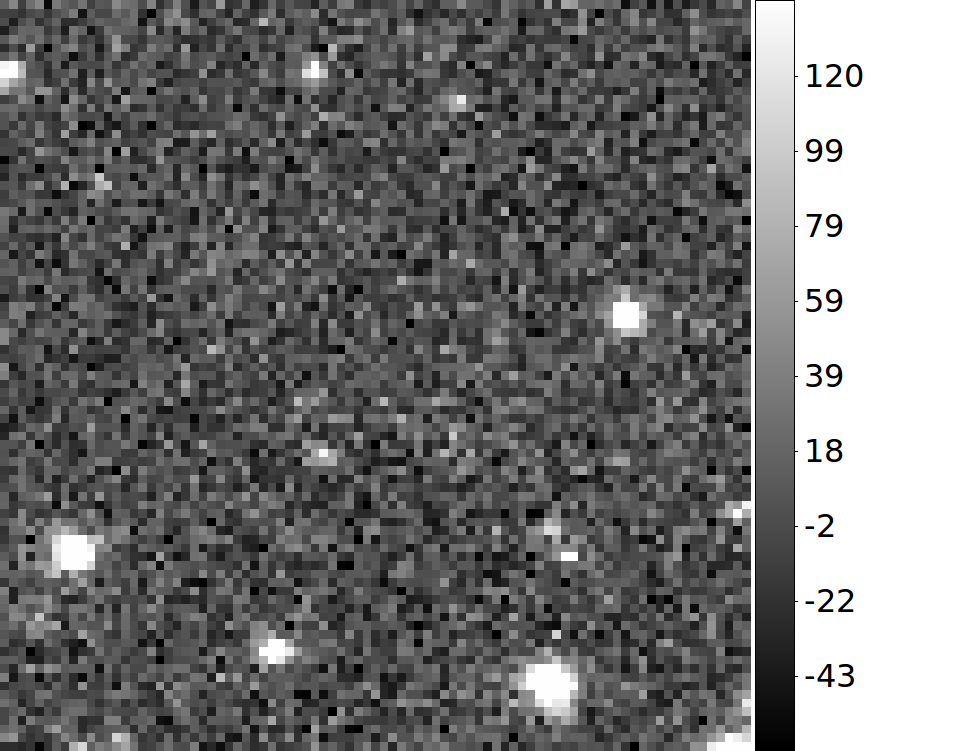}\\
\textit{bad columns of pixel}\\
\includegraphics[trim = 0 0 0 0,clip=true, width=0.45\columnwidth]{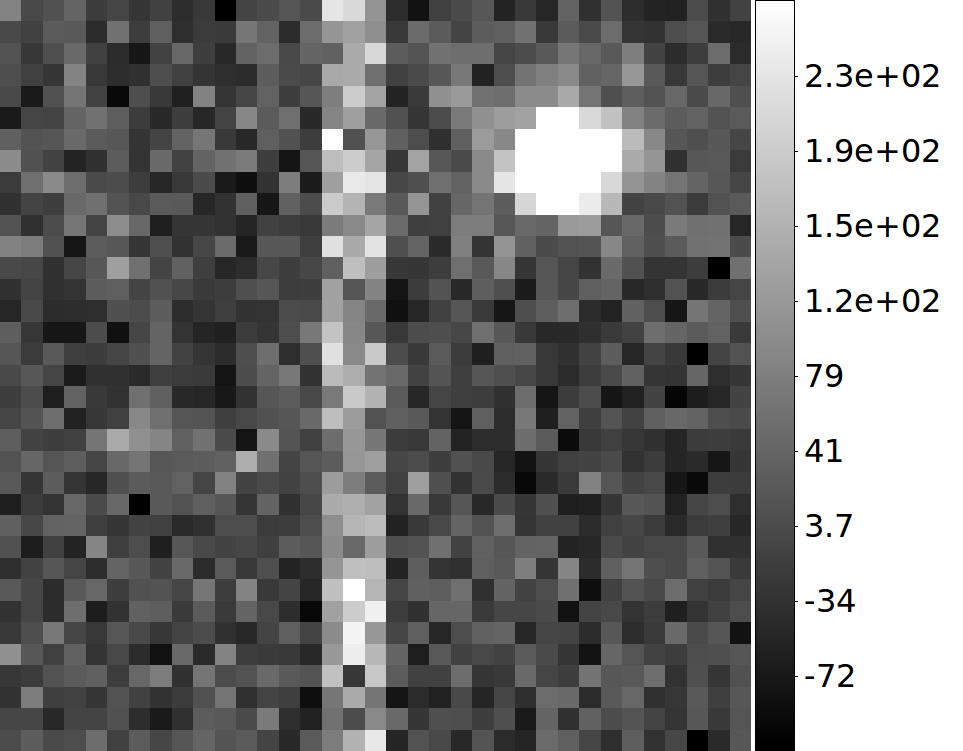}
\includegraphics[trim = 0 0 0 0,clip=true, width=0.45\columnwidth]{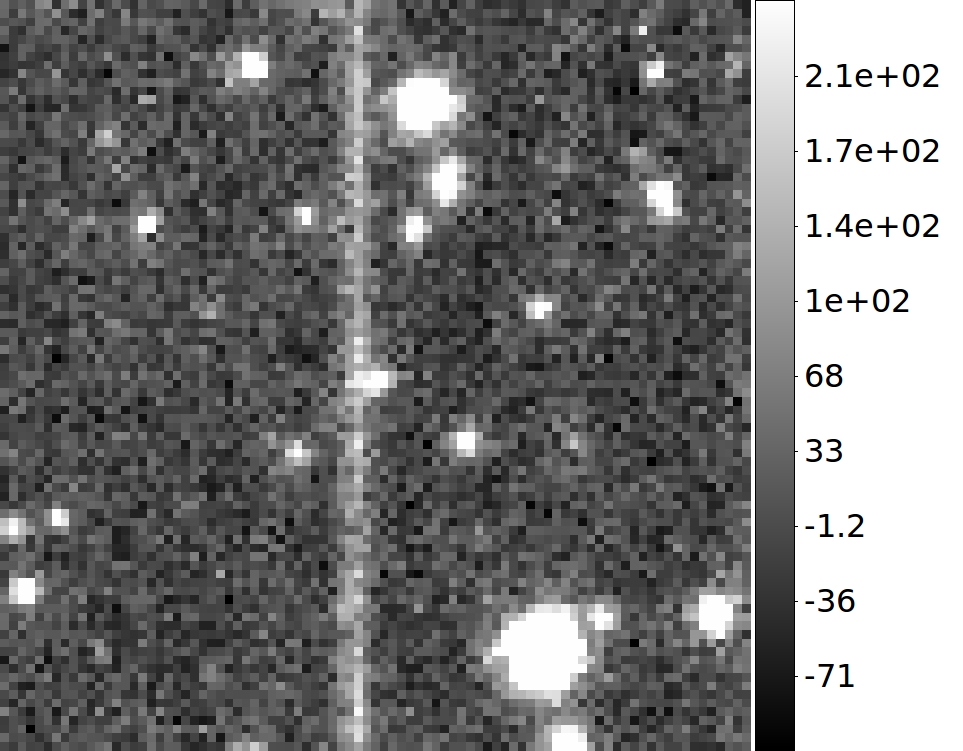}\\
\textit{dark pixels}\\
\includegraphics[trim = 0 0 0 0,clip=true, width=0.45\columnwidth]{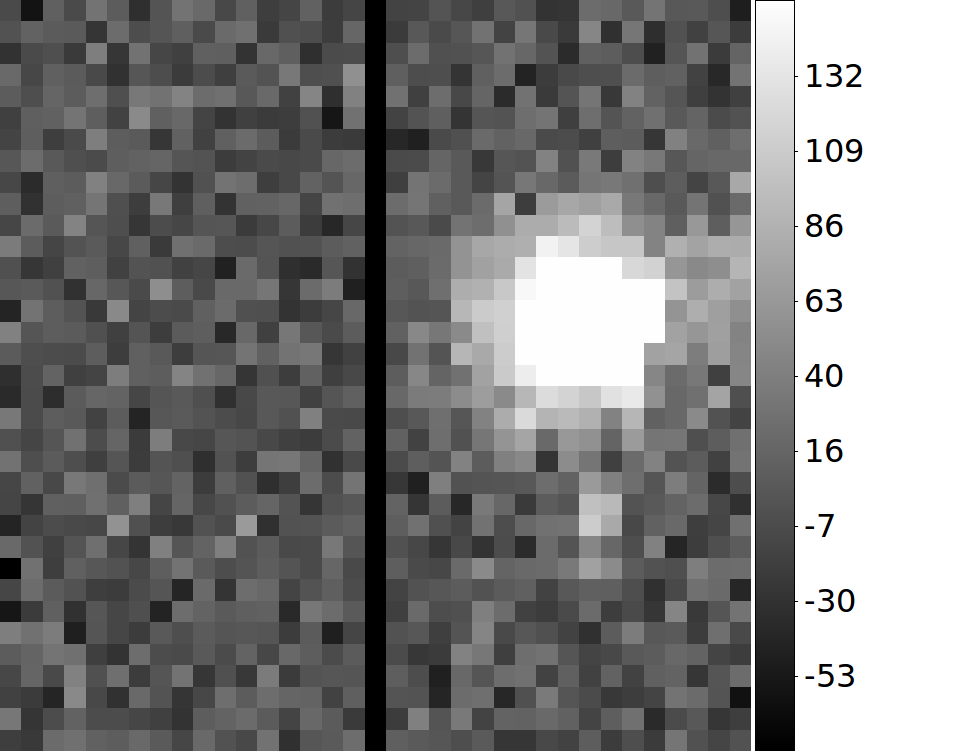}
\includegraphics[trim = 0 0 0 0,clip=true, width=0.45\columnwidth]{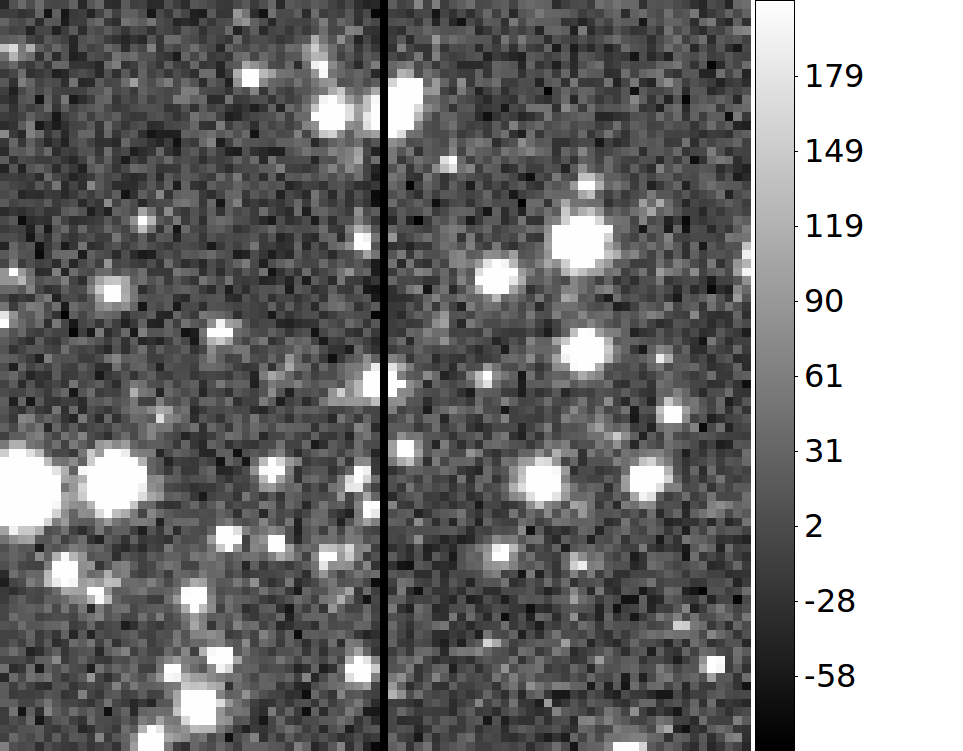}\\
\textit{background with light gradient}\\
\includegraphics[trim = 0 0 0 0,clip=true, width=0.45\columnwidth]{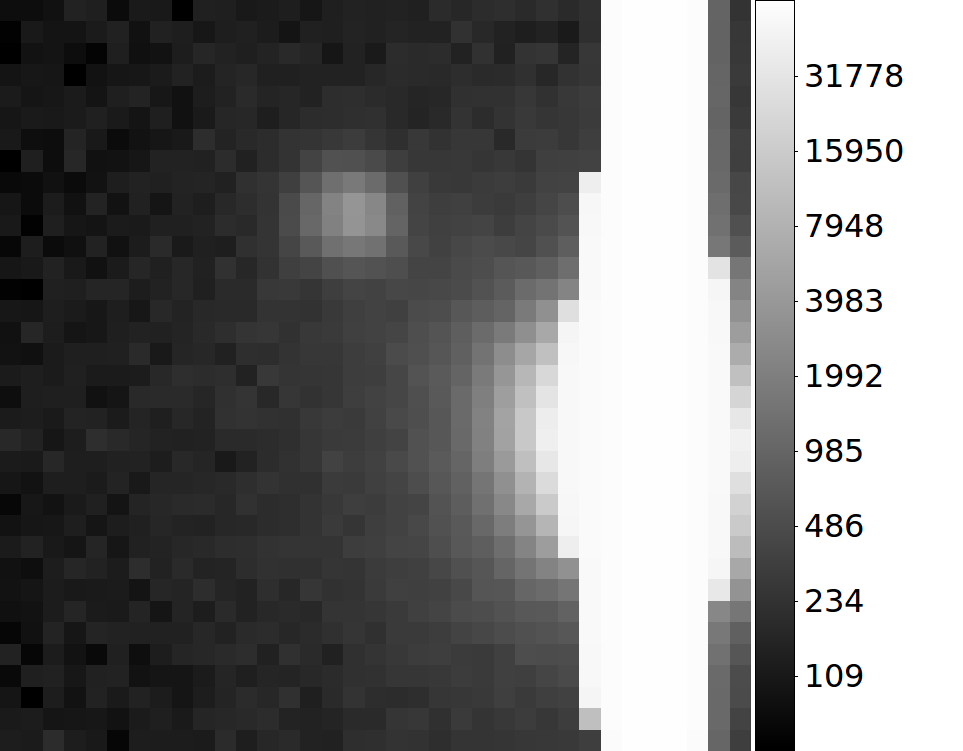}
\includegraphics[trim = 0 0 0 0,clip=true, width=0.45\columnwidth]{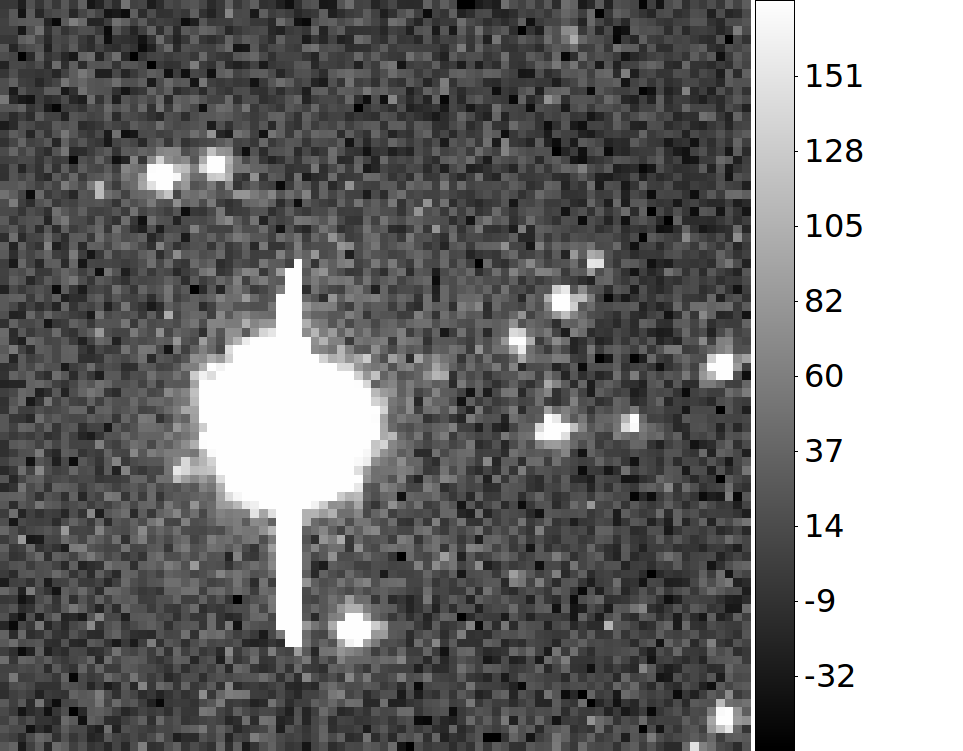}
\caption{The five categories of bogus simulated for the \textit{FOS} training data set (a) compared with (b) the same kind of bogus we indeed observed in the GWAC OT1 finding charts (100 $\times$ 100 pixel-sized).}
\label{fig:FOT_sample}
\end{figure}

For the hot pixels, we actually choose to randomly put a single or a group of hot pixels (2 $\times$ 2 pixels at maximum) at the central position of the sub-images following the pixel intensities we observed from real data. We also add a random increment spanning in [-1;1] pixel to slightly shift the position of the bogus from the center. The bad column of pixels were simulated as an excess of light observed normalised to the pixel intensities we observed for this kind of bogus in real data. Also according to the real data the number of bad columns ranges from 1 to 3 either in the X or Y direction of the image. In figure \ref{fig:FOT_sample}, we compare typical bogus we simulated with the observed ones in the \textit{OT1 candidates} finding charts.

\section{Analysis and results}
\label{sec:results}
The analysis of the classification performance of our CNN is made in two steps: the training to build the keras model and the validation step of the classification procedure on a previously unseen image sample. 
For the training, the \textit{ROS} images are labeled "1" while the \textit{FOS} are labeled "0". We then compared this labeling with the CNN model  probabilistic prediction spanning in the range $P_{CNN}\in[0;1]$. Therefore, a source in a given image is considered as an \textit{FOS} if $P_{CNN}<0.5$ and as an \textit{ROS} if $P_{CNN}\ge0.5$. The mid value 0.5 represents a perfect random guess by the CNN model between the two categories.

\subsection{The training}
\begin{figure}
\centering
\includegraphics[width=0.6\columnwidth]{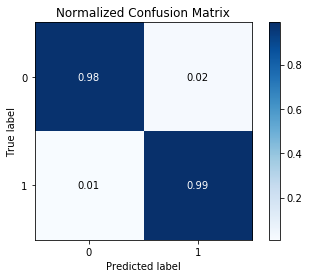}
\caption{The normalized confusion matrix produced after the training of the CNN algorithm on 200 000 simulated images of bogus and real sources. The numbers in each blue square indicate the fraction of the total instances correctly classified as \textit{FP} (top left) and \textit{TP} (bottom right) while in the white squares are shown the mis-classified instances as \textit{TN} (top right) and \textit{FN} (bottom left). For a perfect classifier, the blue squares would indicate "1" while the white squares would indicate "0".}
\label{fig:confu_matrice1}
\end{figure}

\begin{figure}
\centering
\includegraphics[width=0.8\columnwidth]{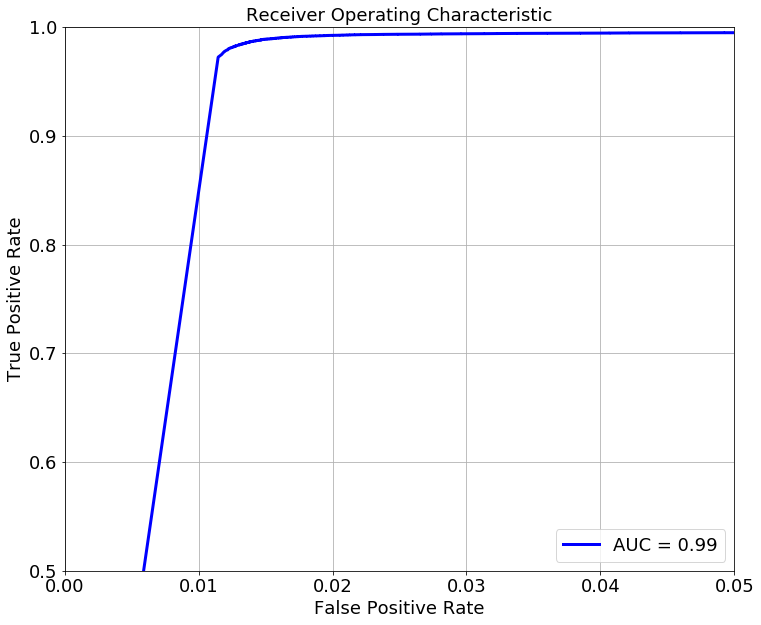}
\caption{The ROC curve of our CNN model applied to the training data set. The AUC value is also indicate at the bottom.}
\label{fig:ROC1}
\end{figure}

We trained our CNN algorithm on the 200 000 simulated images (50\% \textit{ROS}, 50\% \textit{FOS}) making 10 training epochs to build the final \textit{Keras} model. 
Based on the $P_{CNN}$ criteria, we can build the normalized confusion matrix for a quick look of the classification results. The normalized confusion matrices allow to display the fraction of the well classified instances as \textit{TN}, \textit{TP} and the fraction of the mis-classified ones in the \textit{FN} and \textit{FP} categories, as shown in figure \ref{fig:confu_matrice1}. The normalized values of \textit{TN}, \textit{TP}, \textit{FN}, \textit{FP} obey to the following rule:
\begin{equation}
\frac{TN + FN}{N_{FOS}} = 1,~N_{FOS}=10^{5}~~;~~\frac{TP + FP}{N_{ROS}} = 1,~N_{ROS}=10^{5}
\end{equation}
 Based on the training data set, the normalized confusion matrix shows that the CNN algorithm has been well trained to recognize bogus and real sources with classification perfomances close the ideal case where the \textit{TP} and the \textit{FP} instances would be maximized up to a normalize value of "1" while the \textit{FP} and the \textit{FN} would have been minimize to "0".
To better characterize the classification response of our CNN model, we also computed three diagnosis:
\begin{enumerate}
    \item The \textit{receiver operating characteristic} (ROC) curves that display on graph the \textit{True Positive Rate} (TPR) as function of the \textit{False Positive Rate}.
    \item The \textit{Area Under the ROC Curve} (AUC) which corresponds to the integral of the ROC curve $\in[0;1]$. "0" or "1" correspond to an ideal case where 100\% of the instances are mis- or well classified.
    \item The \textit{Accuracy} coefficient (AC) $\in[0;1]$. "0" or "1" correspond to an ideal case where 100\% of the instances are mis- or well classified. :
    \begin{equation}
        AC = \frac{TN+TP}{TN+FP+FN+TP}
    \end{equation}
\end{enumerate}
The ROC curve of the CNN model applied to the training data set also shows that we obtain a very high TPR (close to the maximum value "1") while keeping a extremely low FPR (close to "0"), see figure \ref{fig:ROC1}. This trend is a very convincing proof that a classifier is behaving well as it falsely classifies a very few number of detected events. 
Finally, the corresponding AUC and AC 
are 0.99 and 0.986, respectively, and also point out a very good classification performance of our CNN model. 
All these diagnosis confirm that the architecture of our trained CNN model is well adapted to distinguish bogus form real sources.\\ 
However, while it helps to confirm that the architecture of the CNN is robust enough to perform this kind of classification task, it does not guarantee at all that our CNN model will have the same performance on real GWAC images as our implementation has a limitation. Indeed, we could not simulate all the types of bogus we encounter in the real GWAC images as it would require a too large amount data for the simulation which translates in a higher computational cost and a severe worsening of the simulation complexity. Nevertheless, our FOS simulations and the architecture of our CNN are expected to be generic enough to deal with unseen bogus that may share the same properties than our simulated ones. As an example, we did not simulated any cosmic-ray track in our bogus sample but we simulated some groups of defective pixels that share common properties with those of cosmic-ray tracks (elongated shape with no PSF model or having a very sharp PSF model).\\

\label{sec:res}
\subsection{The validation}

\begin{figure}
\centering
\includegraphics[width=0.8\columnwidth]{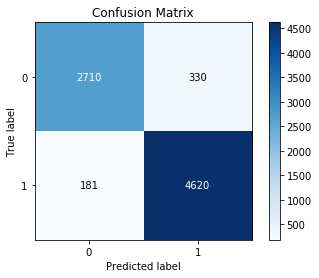}
\includegraphics[width=0.8\columnwidth]{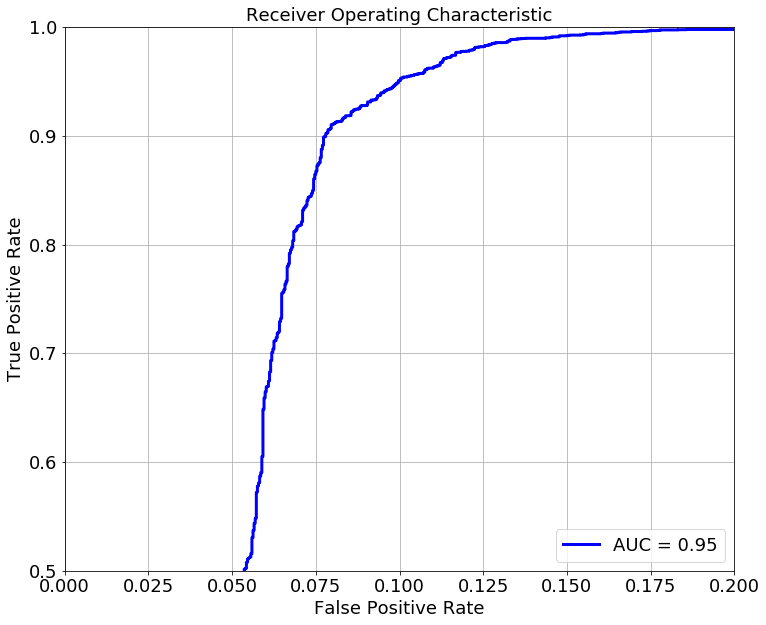}
\caption{The confusion matrix and the ROC curve of our trained CNN model applied on a complete unseen data set of 7841 images of bogus and real astrophysical sources. The classification diagnosis AUC is around 0.95 in good agreement with our scientific classification requirements.}
\label{fig:validation}
\end{figure}

To finally validate the classification performance of our CNN we confront it with a new sample of images representative of the zoo of bogus and real sources the \GWAC pipeline generally detects. Each of those images have been previously labeled by our expert scientists following the same labeling rule described in \ref{sec:CNN_implementation}. In addition to this labeling, each image has been manually classified into representative categories such as real moving objects, hot pixels, 
flaring stars, variable stars, 
bad pixels, dark pixels, incorrectly processed columns of pixels. 
This categorization fits the different groups of bogus used in the simulated training data set and are the most common bogus encountered in GWAC images.
Our validation image sample is finally composed of 7841 images of 1472 objects detected by the GWAC transient search pipeline in 2017 and 2018. The detail of the object distribution into each source category is shown in the table \ref{tab:train_sample}.
\begin{table}
  \begin{center}
    \caption{The different categories of the image sample used to validate the classification performances of our CNN.}
    \label{tab:train_sample}
    \begin{tabular}{l|lll} 
    \hline
      category & $\rm{N_{images}}$& $\rm{N_{object}}$ & typical label \\
      \hline
      \hline
      moving objects & 878  & 29 & 1 (real)\\  
      hot pixels & 1798 & 862 & 0 (bogus) \\
      flaring stars &  102 & 20 & 1 (real)\\  
      variable stars& 3909 & 23 & 1 (real) \\
      bad pixels & 267 & 95 & 0 (bogus) \\
      dark pixels & 333 & 166 & 0 (bogus)\\
      bad pixel columns & 554 & 277 & 0 (bogus) \\
      \hline\hline
      \textbf{Total} & 7841 & 1472 &  --\\
    \end{tabular}
  \end{center}
\end{table}\\
As for the training sample, the probability given by the CNN on each image, P$_{CNN}$, is compared with the image labeling to make the diagnosis of the classification performance.
When we applied our trained CNN model on those images, we finally found that the overall accuracy of the classifier is
AC =0.94 with a very low number of \textit{FN} classification, around 2\% of the total sample. Around only 4\% of the images containing a bogus are misclassified as \textit{ROS} (\textit{FP}) as shown in Figure \ref{fig:validation}. These classification performances are in good agreement with our scientific requirements mentioned in section \ref{sec:classifier} and hence, we consider that our generic deep learning classifier is robust enough to be automatized in the transient detection pipeline as a tool to vet the GWAC optical transient candidates. In table \ref{tab:res_valid_class}, we give more details on the classification performance for each source category used in the validation sample.

\begin{table}
  \begin{center}
    \caption{The results of the CNN classification in the different categories of real/bogus sources tested during the validation step.}
    \label{tab:res_valid_class}
    \begin{tabular}{l|lllll} 
    \hline
      category & TP & TN & FP & FN & AC diag. \\
       & \% &\% &\% &\% & \\
      \hline
      \hline
      moving objects & 0.96 & 0.01 & 0.02 & $<$0.01  & \textbf{0.97}\\  
      hot pixels  & $<$0.01 & 0.93 & 0.06 & $<$0.01  & \textbf{0.94}\\ 
      flaring stars & 0.82 & 0.09 & 0.09 & 0.0  & \textbf{0.91}\\ 
      variable stars& 0.93 & $<$0.01 & 0.06 & $<$0.01  & \textbf{0.94}\\
      bad pixels  & $<$0.01 & 0.94 & 0.05 & 0.0  & \textbf{0.94}\\
      dark pixels & 0.09 & 0.83 & 0.07 & 0.01  & \textbf{0.91}\\
      bad pixel columns & 0.02 & 0.69 & 0.29 & $<$0.01  & \textbf{0.71}\\
    \end{tabular}
  \end{center}
\end{table}
In addition, we also explored the capabilities of our generic CNN model in classifying bogus images that were not included in the simulation of the training sample. We added to our initial validation samples around 1700 images (a data augmentation of $\sim 20$\%) of completely new bogus types such as dust obstructions, suspected \textit{ROS} or low signal-to-noise ratio candidates and a large variety of bad pixels. While the addition of those new bogus make the classification accuracy dropped to AC = 0.91, we found that the performances are still good enough with respect to our scientific requirements. It reinforces the validation test and overall shows how powerful and generic are the CNN algorithms, even with relatively simple layer architectures, in distinguishing any type of bogus from real point-like sources.

\subsection{Analysis of the bogus rejection and false positive detections}
The analysis presented above only considered the classification of the individual images of bogus and real sources. However, the vetting of the OT candidate must also include the time evolution of the source candidates, i.e. taking into account the image time series. Therefore, the rejection of bogus is made on the basis of the evolution of the CNN score across several images. The mean of the CNN probabilities that tracks the stability of the CNN ranking over the image time series is used as a criterion of rejection. Playing on this criterion allows to determine the final rate of \textit{FP} and \textit{FN} the system will tolerate. The current GWAC pipeline is taking a decision on the classification of the candidate after analyzing five consecutive images. For comparison, we used the same numbers of images to take a decision with the CNN. If we have less than five images for a given candidate we computed the mean of P$_{CNN}$ on a minimum of two images. as shown in equation \ref{eq:rejection}, we choose different rejection criteria in order to analyse the evolution of the \textit{FP} and \textit{FN} as function of the strictness of our rejection.
\begin{equation}
\label{eq:rejection}
R=\sum_{i=0}^{N=5}\frac{P_{CNN,i}}{N}\le 5\sigma,~0.997~(3\sigma),~0.95~(2\sigma)~\rm{and }~0.68~(1\sigma)
\end{equation}
A candidate is finally classified as a bogus if it satisfies the rejection criterion otherwise it is classified as a real point-like source.
We applied these criteria to the full validation sample of candidates (1861 candidates including the bogus not simulated in the training data set) and show, in Table \ref{tab:rejection}, the evolution of FN, FP. 
\begin{table}
  \begin{center}
    \caption{The evolution of the \textit{false positives} (\textit{FP}) and \textit{false negatives} (\textit{FN}) as function of the rejection criterion $R \ge 0.99,~0.95,~0.90$.}
    \label{tab:rejection}
    \begin{tabular}{lll} 
    \hline
      R & \textit{FP} & \textit{FN}  \\
       & \% &\% \\
      \hline
      \hline
      R$\le 5\sigma$ & 4.6 & 2.9 \\  
      R$\le 3\sigma$ & 7.2 & 1.7 \\
      R$\le 2\sigma$ & 8.9 & 1.0 \\
      R$\le 1\sigma$ & 10.9 & 0.7 \\
    \end{tabular}
  \end{center}
\end{table}
The goal is to find the good trade-off in the rejection criterion in order to minimize both \textit{FP} and \textit{FN}. A too strict rejection may enhance too much the \textit{FN} while keeping the \textit{FP} very low, i.e.we miss some real events but do not get any fake. On the contrary, a too shallow rejection will go into the opposite direction, i.e. we would keep many bogus by ensuring to keep all the real events.\\
We find that a rejection criterion at R=3$\sigma$ confidence level is finally a good trade-off with less than 2\% of \textit{ROT} loss and about 7\% of false positive detections ($\sim$ 91\% of OT candidate well classified as bogus or real sources). We noticed that the \textit{FN} candidates are actually sources having brightness very close to the detection threshold with a SNR$\le3$ which make them hard to be clearly identified by our CNN algorithm trained on securely detected OT. \\
These results translate into the following scenario. In a typical night, where a hundred of candidates would have been detected, only 7 bogus would have been eventually stored in the \textit{OT1 candidates} list before passing through the series of filter described in section \ref{sec:GWAC_pipeline} and the human/GWAC-F60 telescope vetting step. We find that this number of \textit{false positives} is now easily manageable on a whole night by the GWAC data processing pipeline, the scientist on duty and in addition it will significantly reduce the workload pressure on the GWAC-F60 telescopes scheduler.

\section{Conclusion}
\label{sec:conclusion}
The fast identification and classification of the transient sources are the major challenge to take up for the current and the near upcoming wide field angle facilities dedicated to the time-domain astronomy. In this paper, we have presented a method to distinguish real astrophysical sources from many types of bogus detected by the GWAC survey telescopes (FoV = 25$^\circ$, R$_{lim}$ = 16 in 10s) based on a deep machine learning approach. The machine learning methods are usually easy-to-set-up, cost-effective, time-effective and bring a valuable automated classification procedure to any transient detection pipeline. The first "\textit{True or False}" classification step is now unavoidable to obtain efficient transient search pipeline and quick human reaction to validate the optical transient candidates.
To solve the problem of optical transient vetting in the GWAC images, we used a convolutional neural network classifier trained on computationally-enhanced data relying on the \GWAC database to generate images of real sources and bogus.
\\
The CNN classifier proved to be very efficient in filtering out many types of bogus using a few amount of images for the decision. The final false positive alarm ratio is less than 5\% when it is applied to the individual images. When applying the CNN classifier on the image time series of each OT candidate, we end up with about 7\% of FP optical transient classifications at the level of the \textit{OT1 candidate} sample. The great advantage of our classifier is that it keeps the loss of real OT (FN) as low as 2\% of the total transient candidate sample. This is a key parameter to maintain a high level of transient detection rate every night.\\
Including such classifier tool in the transient detection pipeline of GWAC will significantly lighten the workload pressure of the pipeline itself and the GWAC duty scientists. These performances are in well agreement with the scientific requirements of the GWAC system that aim at detecting and quickly identifying optical transient sources. Therefore, the output CNN score is a precious information for the scientists who will have to take important decisions and actions with respect to any detected OT candidate. Our classifier is generic enough so that a quick configuration of the CNN parameters can also make it usable for other kind of optical facilities.
\\
This work amongst others shows how it is important now for wide field angle telescopes in the time domain astronomy to use such machine learning techniques to deal with huge data flow and big data analysis.

\section*{Acknowledgements}
D.Turpin acknowledges the financial
support from the Chinese Academy of Sciences (CAS) PIFI post-doctoral fellowship program
(program C). D. Turpin is now supported by the CNES Postdoctoral Fellowship at D\'epartement d'astrophysique du CEA-Saclay. M. Ganet acknowledges the financial support of LAL and NAOC. S. Antier is supported by the CNES Postdoctoral Fellowship at Laboratoire AstroParticule et Cosmologie. This work is supported by the National Natural Science Foundation of China (Grant No. 11533003 and 11973055 ) as well as the Strategic Priority Research Program of the Chinese Academy of SciencesGrant No.XDB23040000 and the Strategic Pionner Program on Space Science, Chinese Academy of Sciences, Grant No.XDA15052600.



\bibliographystyle{aa}
\bibliography{aanda}

\label{lastpage}
\end{document}